\documentclass[conference]{IEEEtran}
\IEEEoverridecommandlockouts
\IEEEpubid{
    \makebox[\columnwidth]{
        Proceedings of The 13th International Conference on Computing and Communication Technology (RIVF 2019), Da Nang, Vietnam. Copyright 2019 by the author(s)\hfill
    } 
    \hspace{\columnsep}\makebox[\columnwidth]{ }
}
\usepackage{subcaption}
\usepackage{booktabs}
\usepackage{cite}
\usepackage{amsmath,amssymb,amsfonts}
\usepackage{algorithmic}
\usepackage{graphicx}
\usepackage{textcomp}
\usepackage{xcolor}
\usepackage[bookmarks=false]{hyperref}
\def\BibTeX{{\rm B\kern-.05em{\sc i\kern-.025em b}\kern-.08em
    T\kern-.1667em\lower.7ex\hbox{E}\kern-.125emX}}
\begin{document}

\title{A Simplified Framework for Air Route Clustering Based on ADS-B Data}
\author{%
  \makebox[.5\linewidth]{Quan Duong}\\\textit{ICT Department}\\\textit{John von Neumann Institute}\\Ho Chi Minh, Vietnam\\quan.duong2015@ict.jvn.edu.vn\\
  \and \makebox[.5\linewidth]{Tan Tran}\\\textit{ICT Department}\\\textit{John von Neumann Institute}\\Ho Chi Minh, Vietnam\\tan.tran2015@ict.jvn.edu.vn\\
  \and \makebox[.5\linewidth]{Duc-Thinh Pham}\\\textit{Air Traffic Management Research Institute}\\\textit{School of Mechanical and Aerospace Engineering}\\\textit{Nanyang Technological University}\\Singapore, Singapore \\dtpham@ntu.edu.sg\\
  \and \makebox[.5\linewidth]{An Mai}\\\textit{ICT Department}\\\textit{John von Neumann Institute}\\Ho Chi Minh, Vietnam\\an.mai@jvn.edu.vn
}


\maketitle

\IEEEpubid{\begin{minipage}{\textwidth}\ \\[12pt] \centering
  Proceedings of the $13^{th}$ International Conference on Computing and Communication Technology (RIVF 2019), Danang, Vietnam. ~\copyright~2019~by the author(s)
\end{minipage}}
\IEEEpubidadjcol

\begin{abstract}
The volume of flight traffic gets increasing over the time, which makes the strategic traffic flow management become one of the challenging problems since it requires a lot of computational resources to model entire traffic data. On the other hand, Automatic Dependent Surveillance - Broadcast (ADS-B) technology has been considered as a promising data technology to provide both flight crews and ground control staff the necessary information safely and efficiently about the position and velocity of the airplanes in a specific area. In the attempt to tackle this problem, we presented in this paper a simplified framework that can support to detect the typical air routes between airports based on ADS-B data. Specifically, the flight traffic will be classified into major groups based on similarity measures, which helps to reduce the number of flight paths between airports. As a matter of fact, our framework can be taken into account to reduce practically the computational cost for air flow optimization and evaluate the operational performance. Finally, in order to illustrate the potential applications of our proposed framework, an experiment was performed using ADS-B traffic flight data of three different pairs of airports. The detected typical routes between each couple of airports show promising results by virtue of combining two indices for measuring the clustering performance and incorporating human judgment into the visual inspection.
\end{abstract}


\section{Introduction}\label{section_intro}
Together with the continuously growing need of traveling by aviation, one of the recent most challenging issue arise in air traffic management is flight arrivals delayed in almost all airports in the world. Some of the causes of flight delays or cancellation are maintenance problems related to the aircraft, fueling, inclement weather, etc., and more importantly airline glitches is one of the the top cause of flight delays.
In the attempt to tackle partly the airline glitches issues, people tends to address to a more specific problem of route optimization since it's seen that a non-optimal flight plan route may cause the airline glitches and indirectly cause the flight delay. From this aspect, we propose a simplified framework which leverages the advantage of machine learning to cluster the air route based on trajectory clustering and ADS-B data to support the optimization for the route schedule in air traffic management. Herein, ADS-B data is a kind of data generated by a new surveillance technology, in which the position of an aircraft is determined via satellite navigation and also periodically broadcasted, enabling it to be tracked in real time. ADS-B requires no external input since it depends only on the signal from the navigation system of the aircraft, and hence is more and more promising to provide significant and simplified operational enhancements to military and civilian applications. However, to our best knowledge, there are very few studies until now on leveraging this kind of data in advanced studies from machine learning/AI perspectives for air traffic management. Therefore, there may be a big chance to open up many interesting research on this field of application. 
\IEEEpubidadjcol
From the machine learning perspectives, recently people realizes that there are many rooms for development in trajectory clustering problems due to rapid improvements  in satellites and tracking facilities which make it possible to collect a large amount of trajectory data of moving objects (for e.g., hurricane track data, animal movement data, flight data, etc.). To study the trajectory clustering of moving objects, Gaffney and the team\cite{Gaffney:1999:TCM:312129.312198} 
have proposed a mixture regression model-based trajectory clustering algorithm, in which cluster memberships are determined by using EM algorithm. In \cite{fu2005similarity}, the authors have presented a framework to classify vehicle motion trajectories based on hierarchical clustering; including two steps: trajectories preprocessing and resampling at the first step, then carrying out the trajectories spectral clustering based on similar spatial patterns at the second step. 
To provide a holistic understanding and deep insight into this interesting topic, a comprehensive survey of the development of trajectory clustering was proposed in \cite{bian2018survey}. From this, the authors group the existing trajectory clustering methods into three categories: unsupervised, supervised and semi-supervised algorithms based on machine learning perspectives. They also show the appearance of different trajectory data in many modern intelligent systems for surveillance security, abnormal behavior detection, crowd behavior analysis, and traffic control, which has attracted growing attention. 

Regarding to the data sources, Radar Track data was often used in various works from the literature (see in \cite{CondeRochaMurca2016}, \cite{Eckstein-2009}, \cite{Eerland2016}, and \cite{Gariel_2011}). The other work from Bombelli and the team \cite{Bombelli2017_routeClustering} used the historical data containing the set of Future ATM Concept Evaluation Tool (FACET) Track (TRX) file. And from Yulin Lui \cite{Lui-2017} work, they used the data of the flights' information from FAA Traffic Flow Management System and Aviation System Performance Metric. However, as we know there are recently three publications (\cite{dhief:hal-01592231}, \cite{Junzi_2017}, and \cite{Verbraak_2017}) in the Twelfth USA/Europe Air Traffic Management Research and Development Seminar (ATM2017) shared their studies on ADS-B data. From the work of Junzi and the team \cite{Junzi_2017}, they leveraged the large amount of trajectory data for extracting different aircraft performance parameters and they generated different data mining models corresponding to these parameters. While Dhief and the team \cite{dhief:hal-01592231} makes use of the ADS-B based systems and can prove their effectiveness in oceanic area. The United States have plans for mandating ADS-B Out by 2020 for all airplanes, both air transport and general aviation, and FAA expects to improve the operational performance by using ADS-B based systems. These kinds of system are increasing around the world and applying ADS-B data is an essential requirement for all aircraft operating on the European, Canadian, and Australian in NAS since 2015 \cite{dhief:hal-01592231}.
Besides, there are several investigations about the quality of ADS-B data such as \cite{Barsheshat_2011}, \cite{Busyairah_2013}, \cite{Rekkas_2008}, \cite{Verbraak_2017}. Busyairah and the team \cite{Busyairah_2013} proposed a framework for evaluating ADS-B data in the London Terminal Maneuvering Area, the result showed that 66.7{\%} of aircraft meet the requirement in term of accuracy, integrity, latency, availability, and update rate. Especially, Rekkas and the team \cite{Rekkas_2008} and Barsheshat \cite{Barsheshat_2011} showed similar results, in which their evaluation results considering ADS-B performance are very positive. Barsheshat \cite{Barsheshat_2011} also stated that implementing an ADS-B system provides many benefits, including the reduction for the need of maintaining and/or upgrading radar infrastructure.

For the behind techniques, an extensive comparative study of cluster validity indices has been carried out by Arbelaitz and the team \cite{Arbelaitz_2013}. They experimented with 30 different cluster validity indices with the goal is to choose the best index for each individual application. From their work, 
the authors also recommend to employ several indices at the same time to receive the robust results. Regarding to trajectory clustering evaluation, most of the works (\cite{Eckstein-2009}, \cite{Gariel_2011},\cite{Lui-2017}, and \cite{Adria_2015}) preferred to use Silhouette score alone for choosing the best number of clusters. Another combination from the work of Bombelli and the team \cite{Bombelli2017_routeClustering} leveraged three different indices which is average Silhouette index, Davies-Boulding, and Dun index. And the other work from Maraya and the team \cite{CondeRochaMurca2016} combined Silhouette score and Davies-Bouldin (DB) score.

Inspired by the above results, we propose to leverage the ideas from these analysis in a combination with a simplified framework, in order to adapt for the case of ADS-B data in air route clustering problem. The paper is organized as follows: In Section \ref{section_intro}, we discussed about the motivation and previous works related to this paper. Then the Section \ref{section_proposed_fw} presents our proposed framework, in which the considered ADS-B data and appropriate techniques involving in the step of data preprocessing will be covered respectively in Section \ref{sec_data} and \ref{sec_data_process}. The experimental design and corresponding results taking into account real ADS-B data will be placed in Section \ref{sec_result}. Finally, we come to a conclusion and future works  in Section \ref{sec_conclusion}.

\section{A PROPOSED FRAMEWORK}\label{section_proposed_fw}

\begin{figure*}
\centering
\includegraphics[width=1.\textwidth]{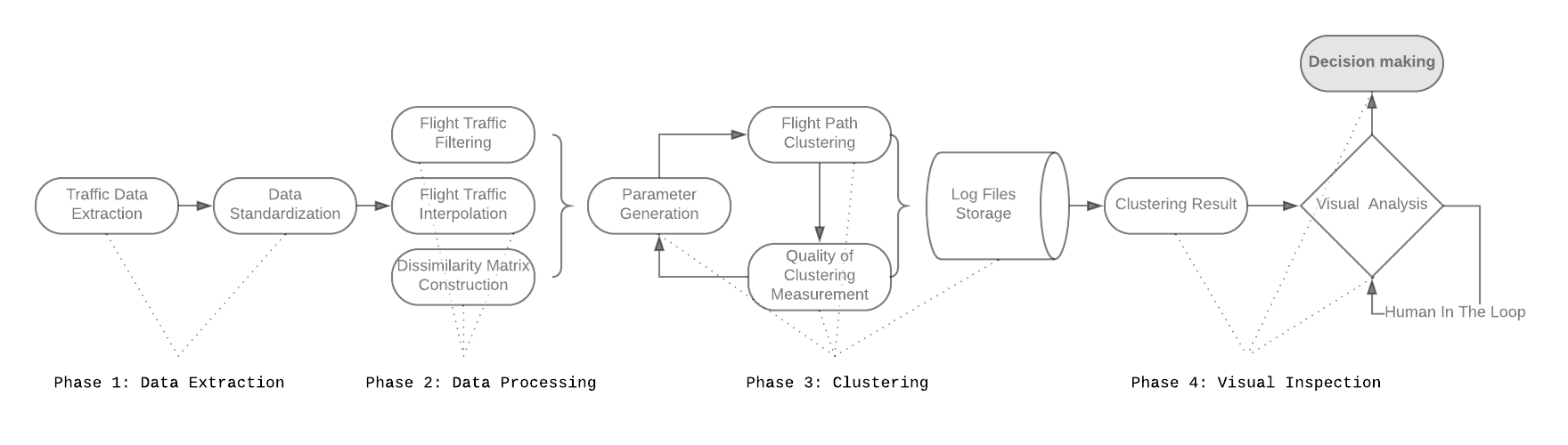}
\caption[Trajectories Clustering Flow]{The detail flow of our proposed framework for air route clustering based on ADS-B data}
\label{fig:our_proposed_flow}. 
\end{figure*}

\begin{table}
\caption{The Benchmark Methodologies}
\centering
\label{table:benchmark_sum}
\begin{tabular}{l l}
\toprule
\multicolumn{1}{l}{Step} & \multicolumn{1}{l}{Methods}\\ 
\midrule
Trajectory data & Radar track\\
Distance between curve & Fr{\'e}chet distance\\
Clustering & DBSCAN\\
Decide Number of Cluster & Silhouette Score\\
\bottomrule
\end{tabular}
\end{table}

\begin{table}
\caption{Trajectories Flow Clustering Description}
\label{table:our_methods}
\begin{tabular}{p{3cm} p{5cm}}
\toprule
\multicolumn{1}{c}{Step} & \multicolumn{1}{c}{Description}\\ 
\midrule
1.1 Traffic Data Extraction & Received raw traffic data in JSON format\\
1.2 Data Standardization & Group traffic data to identified flights and convert to CSV format\\
2.1 Flight Traffic Filtering & Filter out the unwanted airports, remove flights that land or depart outside of terminal, and drop the duplicated data points \\
2.2 Flight Traffic Interpolation & Convert individual flight's tracks into vectors of equal length. Apply the cubic spline \cite{Dierckx_smoothing_1982} to interpolate the spaced data points\\
2.3 Dissimilarity Matrix Construction & Measure the dissimilarities between flight paths with the same number of features by computing Fr{\'e}chet distance\\
3.1 Parameters Generation & The clustering algorithm replied on two input parameters are $\varepsilon$ (a distance threshold) and MinPts (a minimum number of points). 100 values of $\varepsilon$ were generated and MinPts was retrieved by observations \\
3.2 Flight Path Clustering & Apply a density-based clustering algorithm called DBSCAN with defined MinPts and $\varepsilon$\\
3.3 Quality of Clusters Measurement & Silhouette and Davies-Bouldin indices were combined for measuring the clustering performance\\
3.4 Log files Storage & Record all the performances and were produced for each parameter\\
4.1 Clustering Result & Visualize the detected clusters along with original and interpolated trajectories\\
4.2 Visual Analysis & Analyze the clustering outcome with the recommended ranking from step 3.3, 3.4, and incorporate with human judgment \\
4.3 Decision Making & Make the decision by virtue of domain expert\\
\bottomrule
\end{tabular}
\end{table}

For the main technical part of this paper, we perform the trajectory clustering based on DBSCAN \cite{Ester:1996:DAD:3001460.3001507} approach and ADS-B data, then combine two different evaluation indices (Silhouette and DB) for the clustering outputs, which were used to support the visual inspection step. The detail flow of our framework are shown in Figure \ref{fig:our_proposed_flow}, in which the flow composes four different phases including Data Extraction, Data Processing, Clustering, and Visual Inspection. Additionally, the detail description of involved techniques for each step in the framework can be found in the Table \ref{table:our_methods}. Finally, for a comparison, all experiments are carried out by using the same data, on specific dates, and for particular pairs of origin and destination airports. For a benchmark, we re-implement the route clustering framework from Adria \cite{Adria_2015}, which is summarized the used methodologies in the Table \ref{table:benchmark_sum}, considering the same assumption for both of their system and ours.

\subsection{Data Description}\label{sec_data}
\begin{figure*}
\begin{subfigure}{\linewidth}
  \includegraphics[width=.32\linewidth]{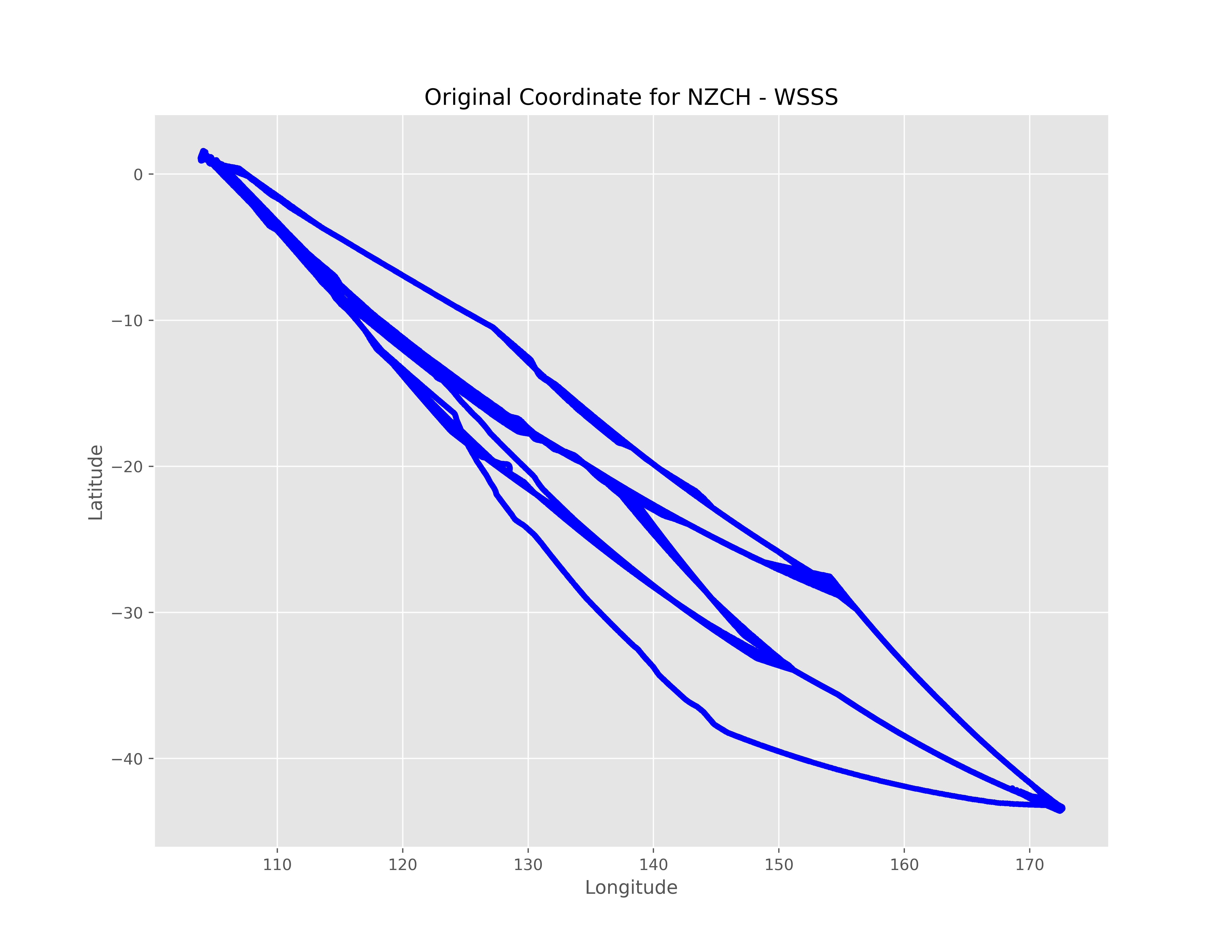}\hfill
  \includegraphics[width=.32\linewidth]{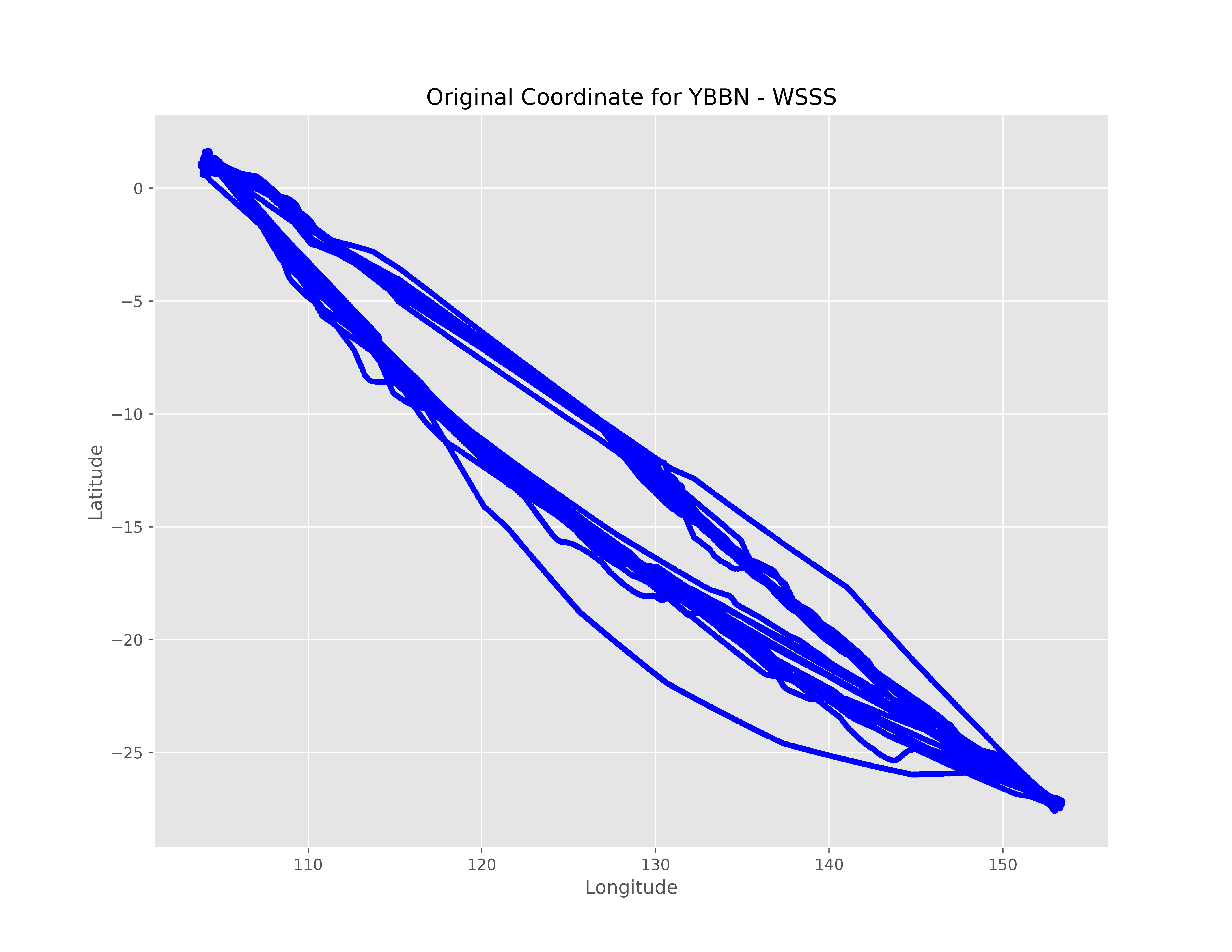}\hfill
  \includegraphics[width=.32\linewidth]{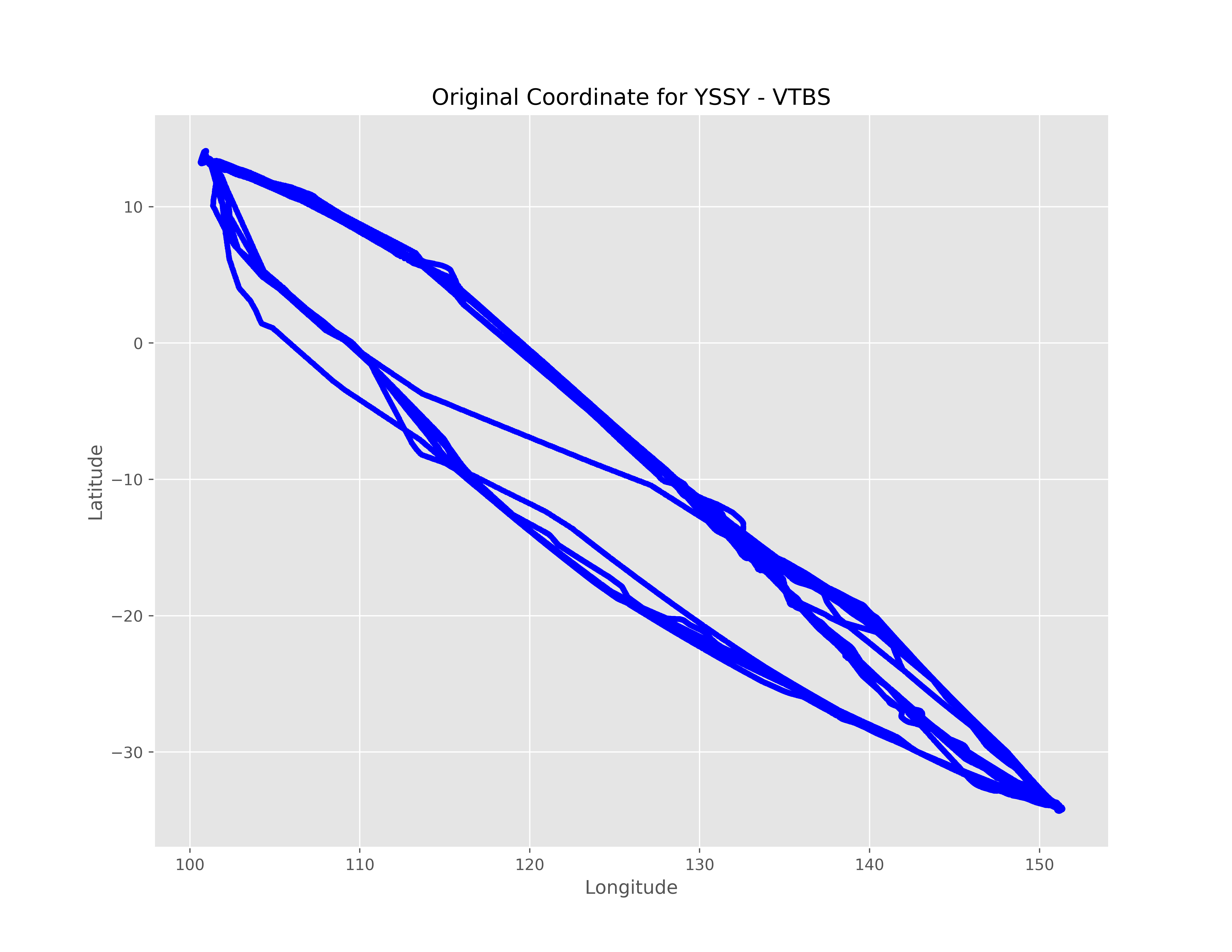}
  \caption{Original Trajectories}
  \label{fig:bm_original_trajectory}
  \end{subfigure}\par\medskip
  \begin{subfigure}{\linewidth}
  \includegraphics[width=.32\linewidth]{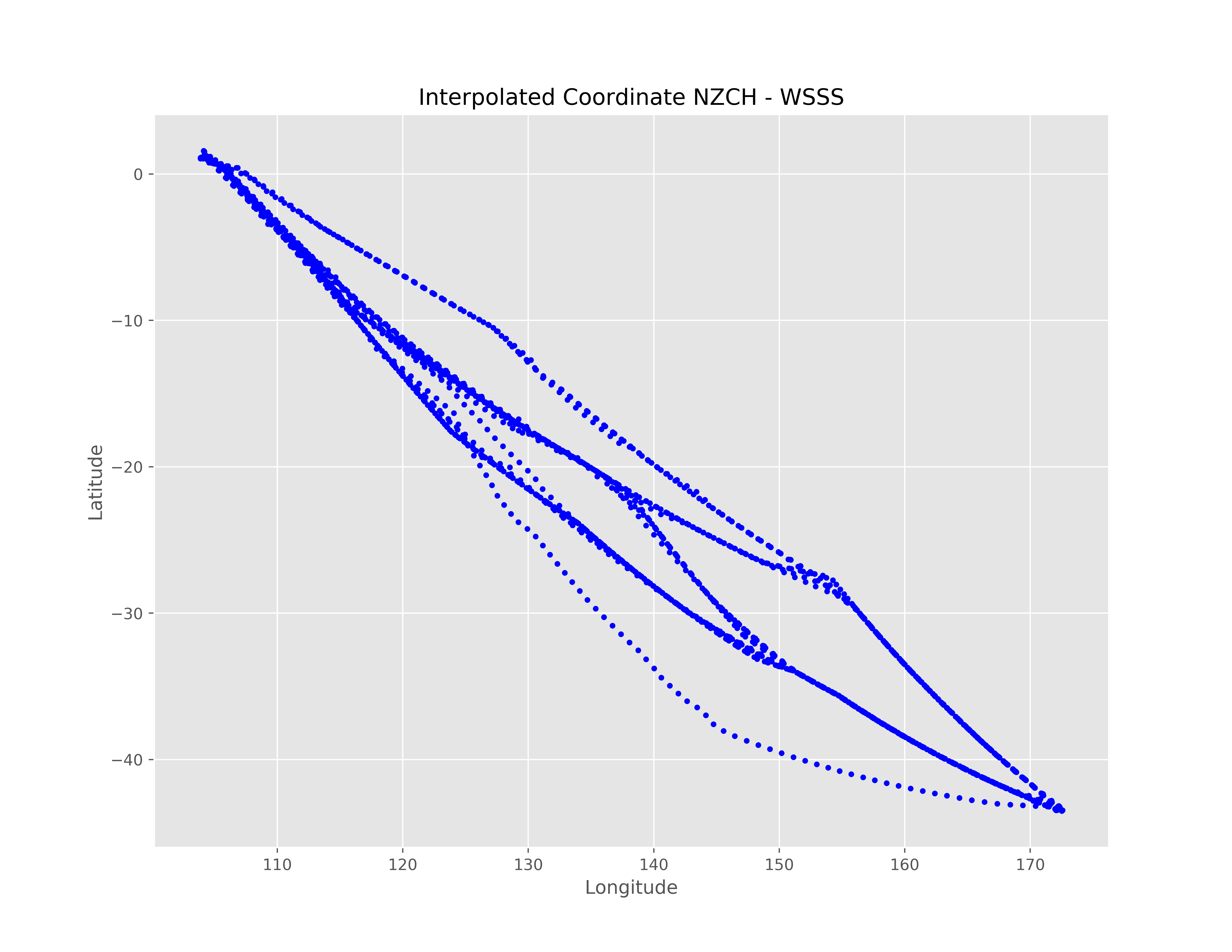}\hfill
  \includegraphics[width=.32\linewidth]{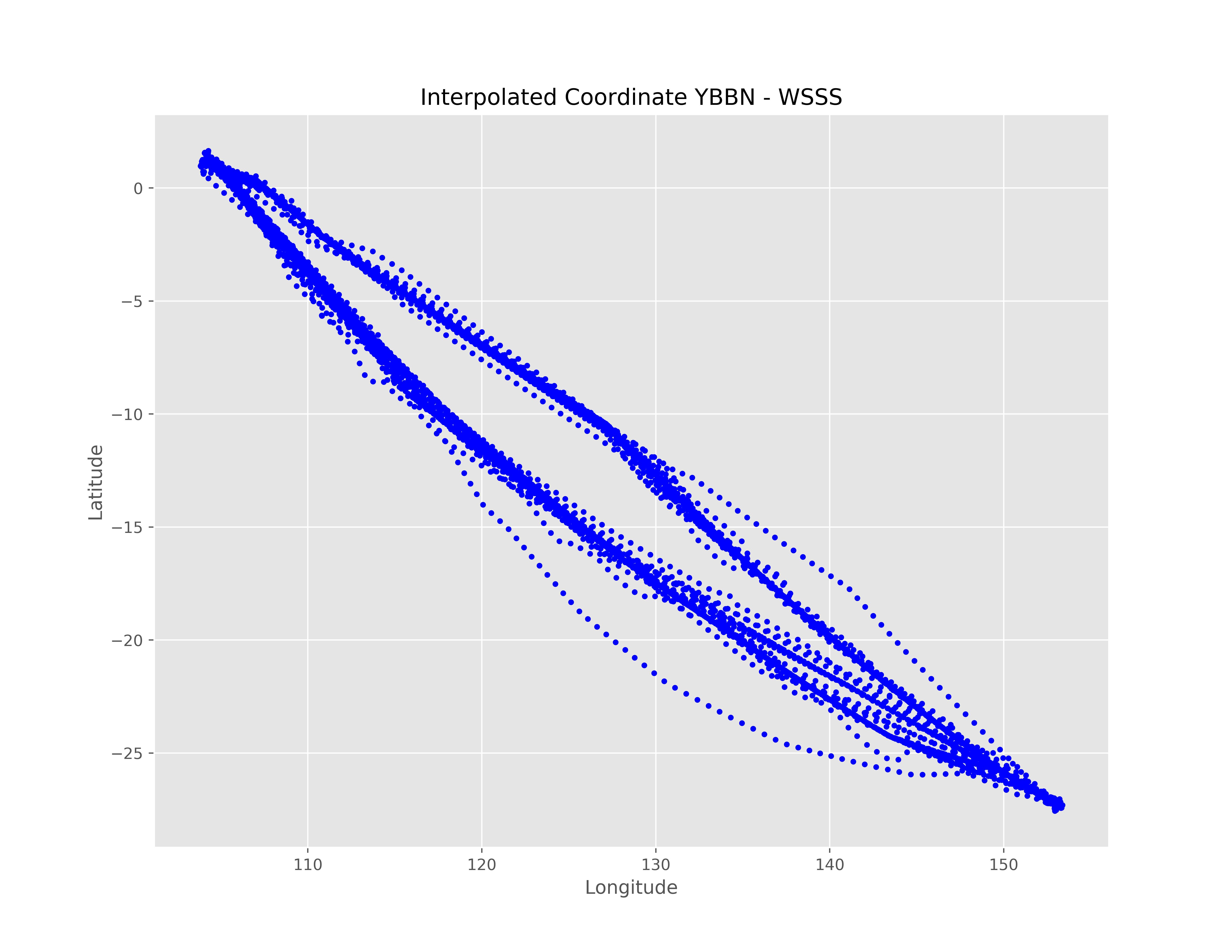}\hfill
  \includegraphics[width=.32\linewidth]{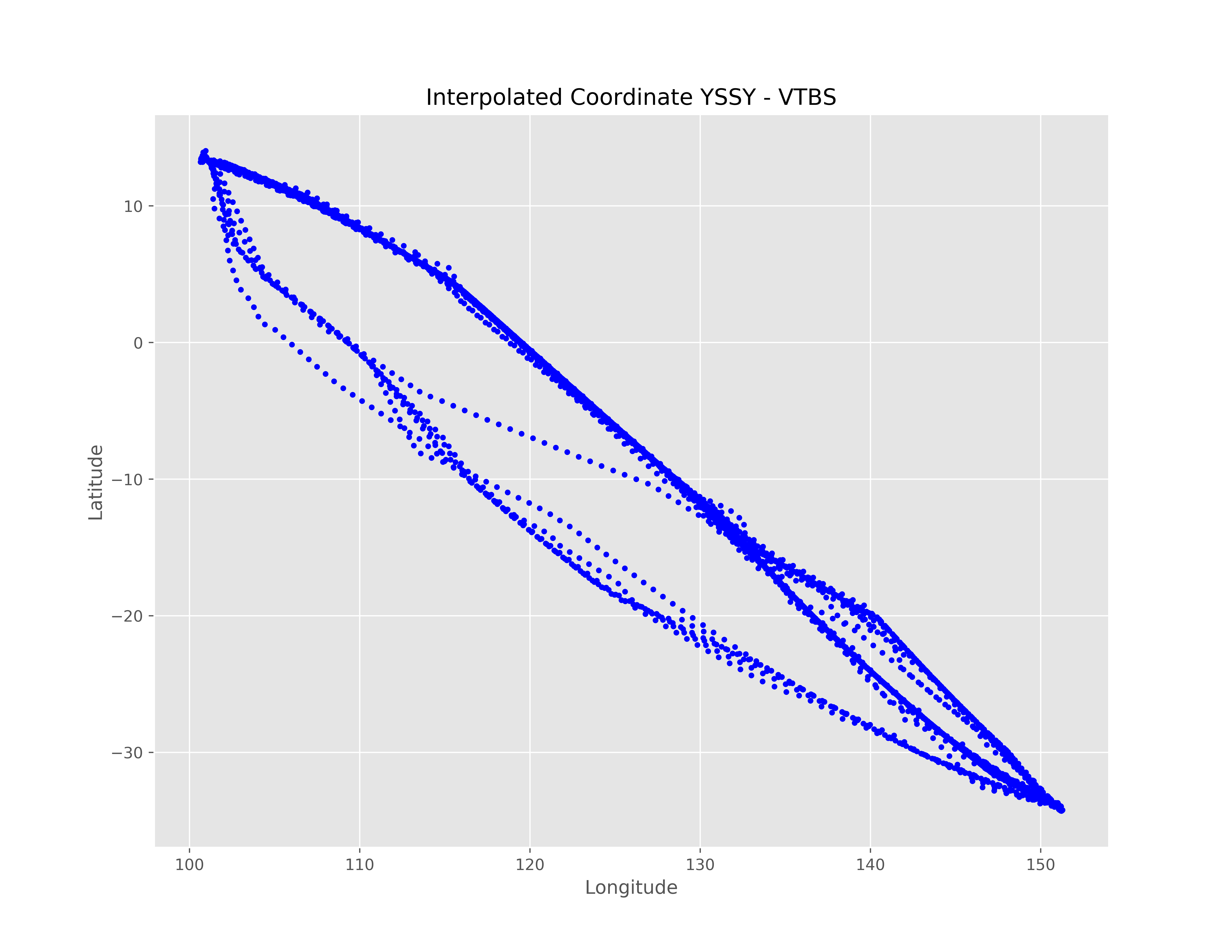}
  \caption{Interpolated Trajectories}
  \label{fig:bm_interpolated}
  \end{subfigure}\par\medskip
  \caption{The Visualization of Trajectories in Data Processing Step performed by our framework. Figures \ref{fig:bm_original_trajectory} is done by using original trajectories up to 2057 spaced points. Figures \ref{fig:bm_interpolated} showed the interpolated trajectories with only 100 spaced points}
\label{fig:data_processing_results}
\end{figure*}

\begin{figure*}
  \begin{subfigure}{\linewidth}
  \includegraphics[width=.32\linewidth]{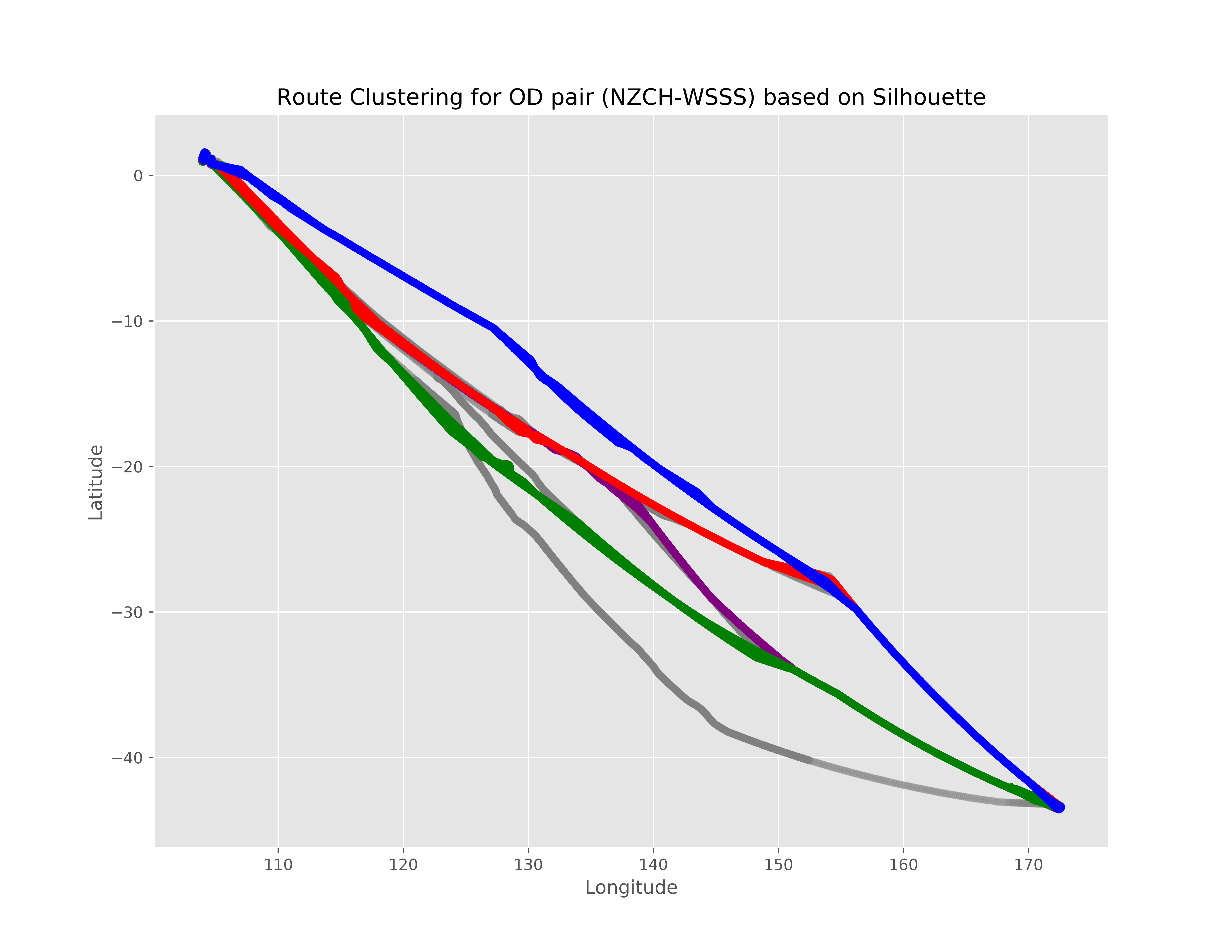}\hfill
  \includegraphics[width=.32\linewidth]{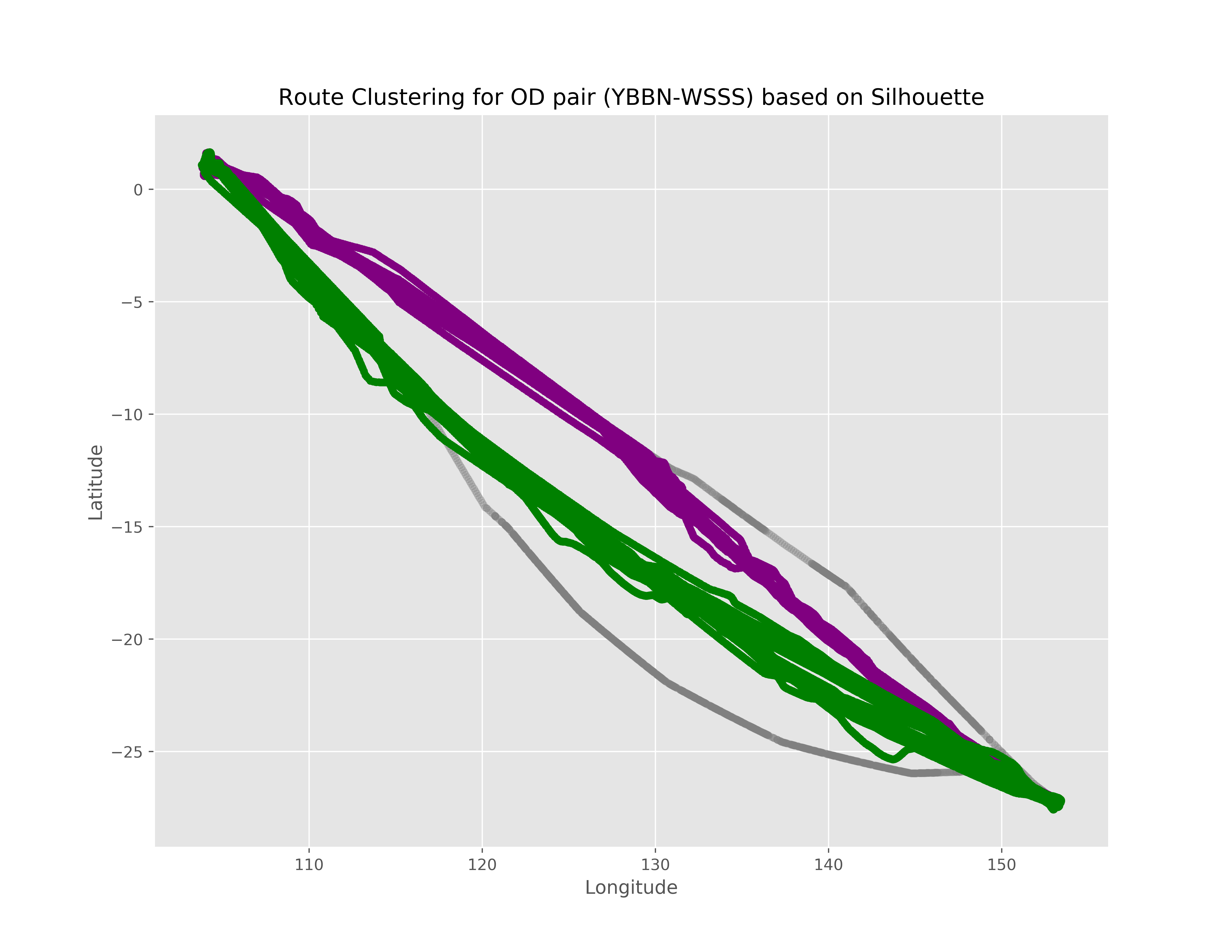}\hfill
  \includegraphics[width=.32\linewidth]{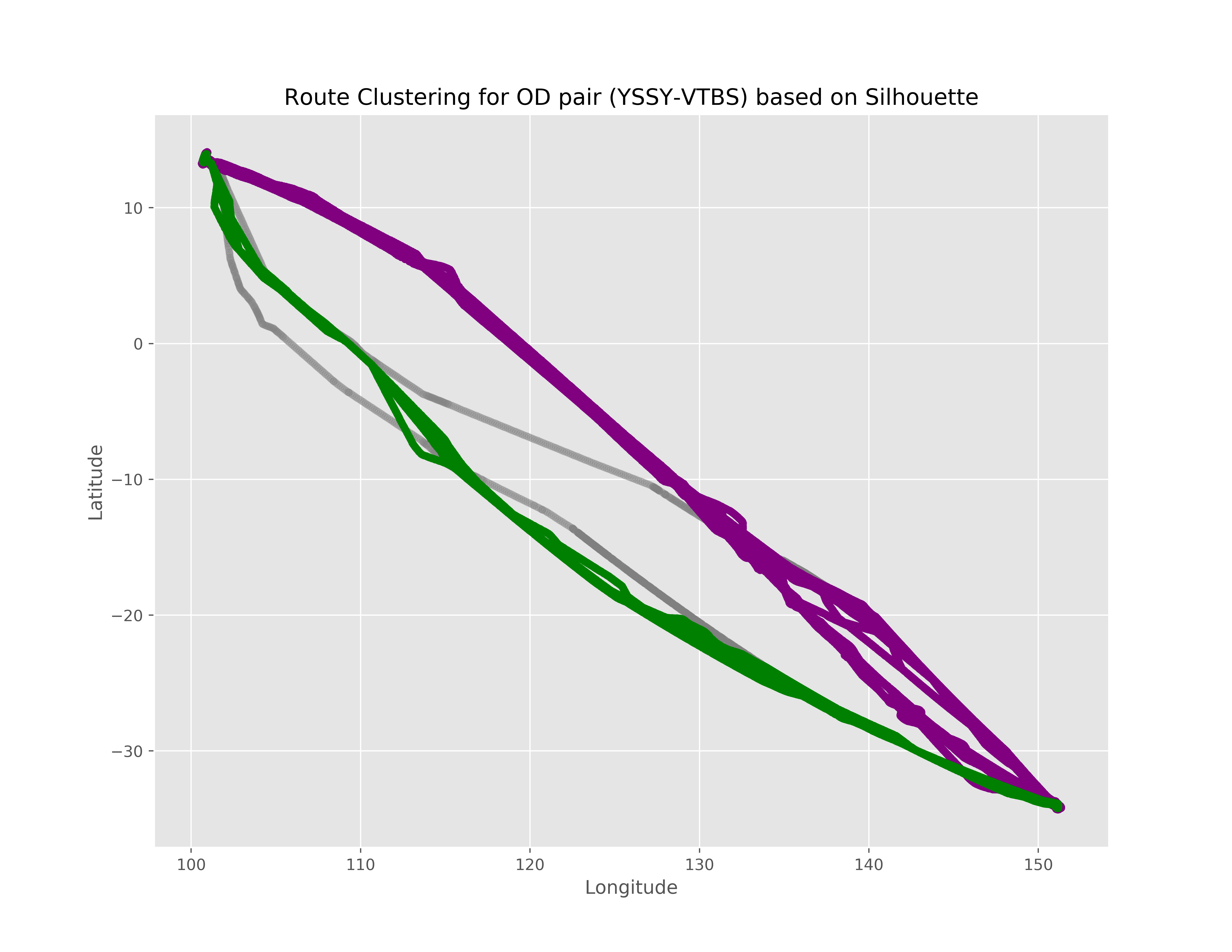}
  \end{subfigure}\par\medskip
  \caption{The result from benchmark method discussed in \cite{Adria_2015}, adapted to ADS-B data. Here, each color represents for each cluster while the grey color is always used to represent for detected outlier trajectories}
\label{fig:benchmark_results}
\end{figure*}


\begin{figure}
  \begin{subfigure}{\linewidth}
  \includegraphics[width=1.\linewidth]{YSSY_VTBS_silhouette_0_920737511764}
  \caption{Best air route clustering result based on single Silhouette index}
  \label{rs_silhouette}
  \end{subfigure}\par\medskip
  \begin{subfigure}{\linewidth}
  \includegraphics[width=1.\linewidth]{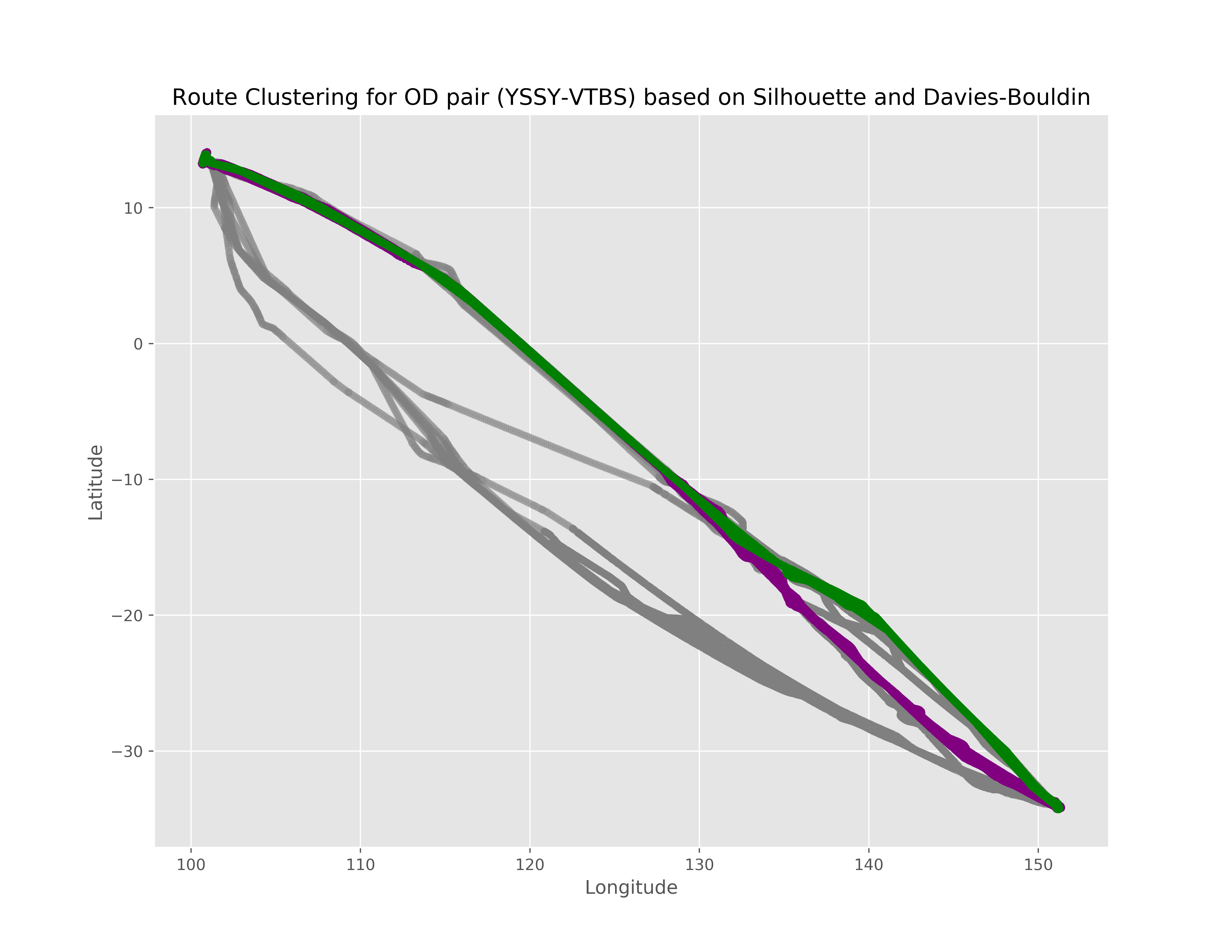}
  \caption{Best air route clustering result based on Silhouette-DB index}
  \label{rs_db_silhouette}
  \end{subfigure}\par\medskip
  \begin{subfigure}{\linewidth}
  \includegraphics[width=1.\linewidth]{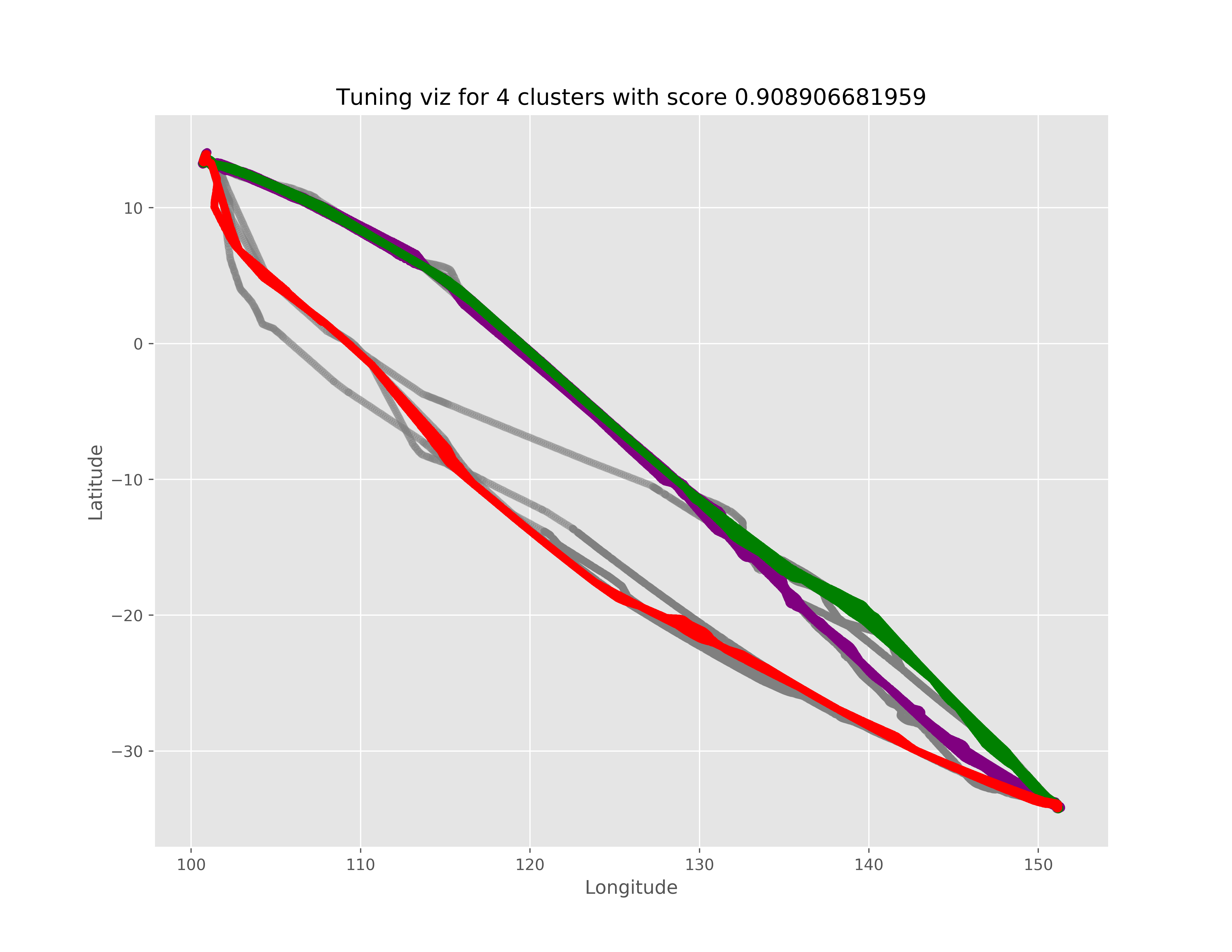}
  \caption{Best air route clustering result based on Visual Inspection step}
  \label{rs_visual_inspection}
  \end{subfigure}\par\medskip
  \caption{Our framework's result for the detected clusters from different scenarios, in case of Sydney-Suvarnabhumi airports}
\label{fig:our_framework_results}
\end{figure}

All flights' trajectories data related to this paper are crawled from \textit{FlightAware}\footnote{\url{https://flightaware.com}}, the company operates a website and mobile application which offers free flight tracking services for both private and commercial air traffic. FlightAware produces FlightFeeder which is known as a network of ADS-B receiver. The message consisting of the ADS-B data is received from airplane transponders directly via a small antenna. Then the ADS-B data is extracted from the received messages and uploaded to FlightAware's server over any available internet connection.
For carrying out the analysis, we are going to employ only the flight trajectory data from three different Origin and Destination pairs of airport, which are shown clearly in the Table \ref{table:od_pair_sum}. The chosen flights for this analysis are collected for entire January 2017.

Here, the whole process of data generation involves of satellites, transmitters, and receivers. The airplane determines its own position and velocity by communicating with satellites via GNSS (global navigation satellite system) or GPS receivers. Then it broadcasts the its position and velocity via ADS-B Out to other airplanes and ground ADS-B receivers. Ultimately, the receivers around the world send this information to FlightAware's servers over a real-time connection.

\subsection{Data Processing}\label{sec_data_process}
After receiving raw traffic data in JSON format, firstly the data standardization steps were employed by grouping traffic flights and converting them to CSV format, that can be consumed for other tasks such as EDA (exploratory data analysis), data processing, and data clustering. Secondly, in order to make sure the clustering algorithm consuming qualified data, the flight traffic filtering step was executed, including filter out the unwanted pairs of origin and destination terminals, remove flights that land or depart at the coordinates outside of terminals, and drop the duplicated data points. The Figure \ref{fig:bm_original_trajectory} show visually the original trajectories after filtering step. Finally, by virtue of flight traffic interpolation, each individual flight's trajectories will be converted into vector of equal length. We then apply the cubic spline interpolation \cite{Dierckx_smoothing_1982} to the spaced data points with the assumption that the spatial and temporal dimensions are used here in the phase. The Figure \ref{fig:bm_interpolated} then visualize the interpolated trajectories, which are able to represent the entire flight's path with a pre-defined number of spaced points.

\section{Experimental Results}\label{sec_result}
This section intents to show the major results obtained from reimplementing the benchmark method from Aria \cite{Adria_2015} in case of our adaptation for ADS-B data, and incorporating our framework's application with human intervention in the visual inspection step. Specifically, a density-based clustering algorithm called DBSCAN is taken into account in the Flight Path Clustering step. The DBSCAN algorithm receives two primary parameters (\textit{MinPts} and \textit{$\epsilon$}) to cluster the trajectory data, where the \textit{MinPts} is the minimum number of data points belong to a cluster determined by the actual observation, and \textit{$\epsilon$} is a distance threshold. In this experimental setting, we generated 100 values for $\epsilon$ combined with the real observed \textit{MinPts} value for running the DBSCAN algorithm. The best parameters is determined by using two evaluation stages, \textit{quantitative} metrics including Silhouette and DB (known as cluster validity indices), and a \textit{qualitative} validation in visual inspection phase. 
\begin{table}
\caption{Origin Destination Pair Summary}
\centering
\label{table:od_pair_sum}
\begin{tabular}{l l l}
\toprule
\multicolumn{1}{l}{Codename} & \multicolumn{1}{l}{Airports} & \multicolumn{1}{l}{Cities}\\
\midrule
YBBN-WSSS & Brisbane-Changi & Brisbane-Singapore\\
YSSY-VTBS & Sydney-Suvarnabhumi & Sydney-Bangkok\\
NZCH-WSSS & Christchurch-Changi & Christchurch-Singapore\\
\bottomrule
\end{tabular}
\end{table}

In Figure \ref{fig:data_processing_results}, we make a visual comparison of the flight paths based on original trajectories and interpolated trajectories from three different pairs of airports (see Table \ref{table:od_pair_sum}), in which you can see that only using 100 spaced points we could describe the whole flight paths between two airports. Next, the Figure \ref{fig:benchmark_results} shows the final clusters of flight paths for three different pairs of airports, in which the highest Silhouette score is selected as best result. Here, each color represents for each cluster while the grey color is always used to represent for detected outlier trajectories in this analysis. In our framework, to obtain more robust results we combine two different indices for measuring the performance of clustering algorithm. In particular, the Silhouette \cite{Pete-1987} and Davies-Bouldin (DB) \cite{Davies-1979} indices are employed altogether to select the best clustering result, in the event of these two indices all agree in choosing the most appropriate route clusters that supports strongly for decision making. On the other hand, when the Silhouette index and combined Silhouette-DB index show different rankings (see in Figure \ref{rs_silhouette} and Figure \ref{rs_db_silhouette}), the visual inspection step will be activated to help the user makes the final decision on air route clustering (see Figure \ref{rs_visual_inspection}).

\begin{figure}
  \begin{subfigure}{\linewidth}
  \includegraphics[width=1.\linewidth]{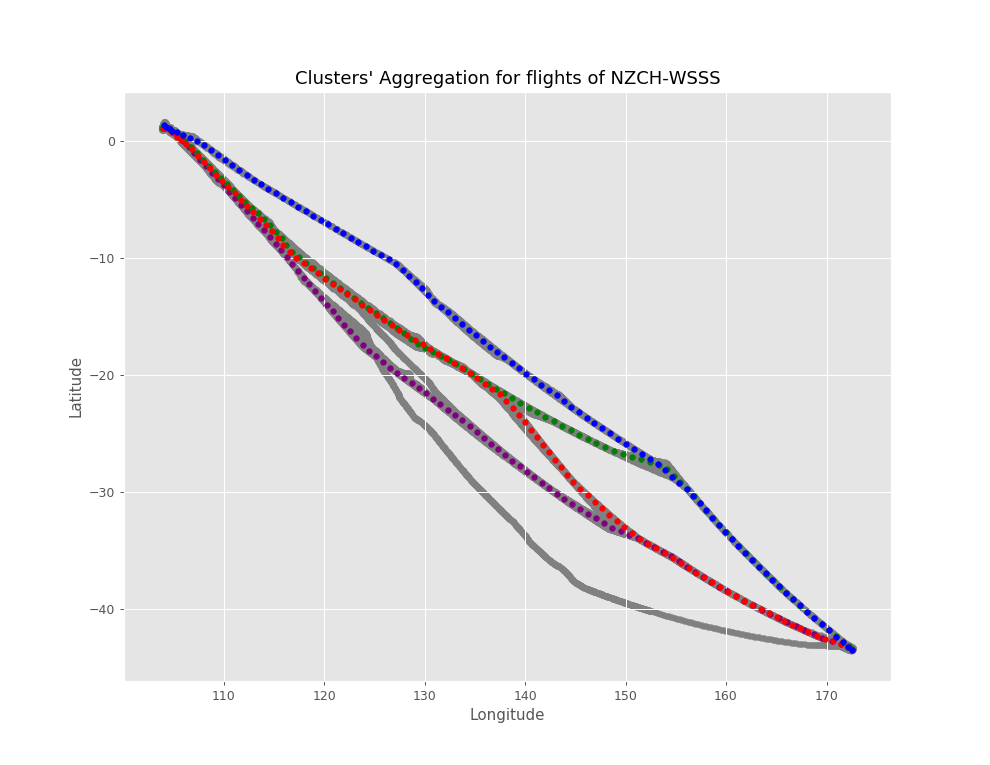}
  \caption{Aggregated Routes in case of Christchurch-Changi airports}
  \label{fig:nzch_wsss_agg}
  \end{subfigure}\par\medskip
  \begin{subfigure}{\linewidth}
  \includegraphics[width=1.\linewidth]{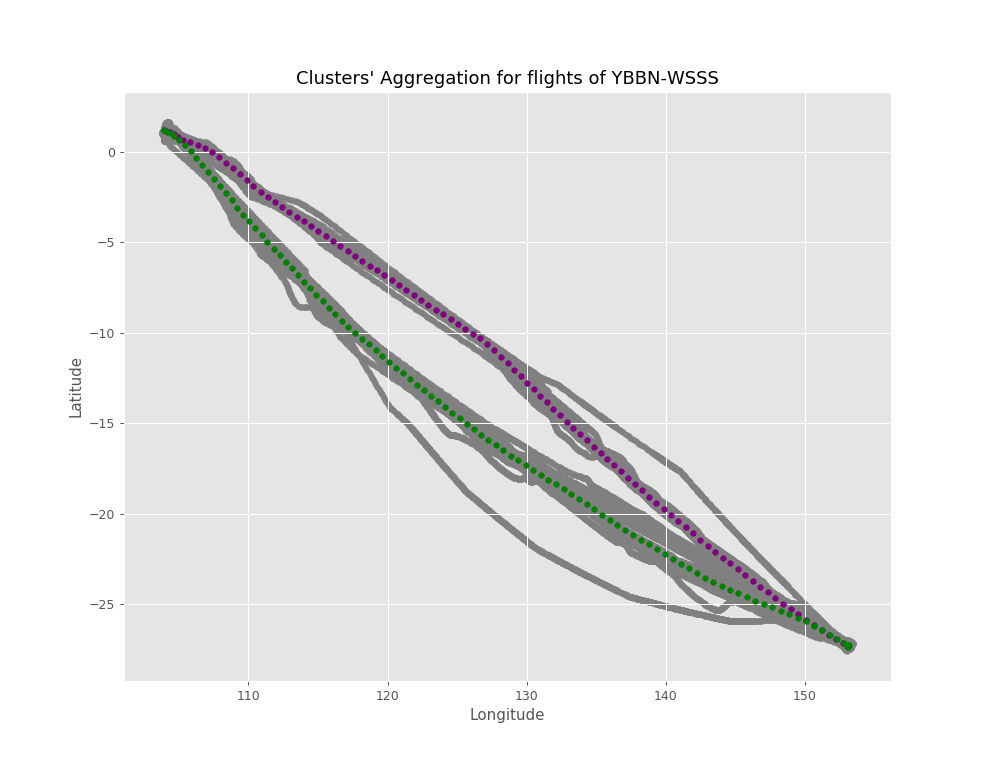}
  \caption{Aggregated Routes in case of Brisbane-Changi airports}
  \label{fig:ybbn_wsss_agg}
  \end{subfigure}\par\medskip
  \begin{subfigure}{\linewidth}
  \includegraphics[width=1.\linewidth]{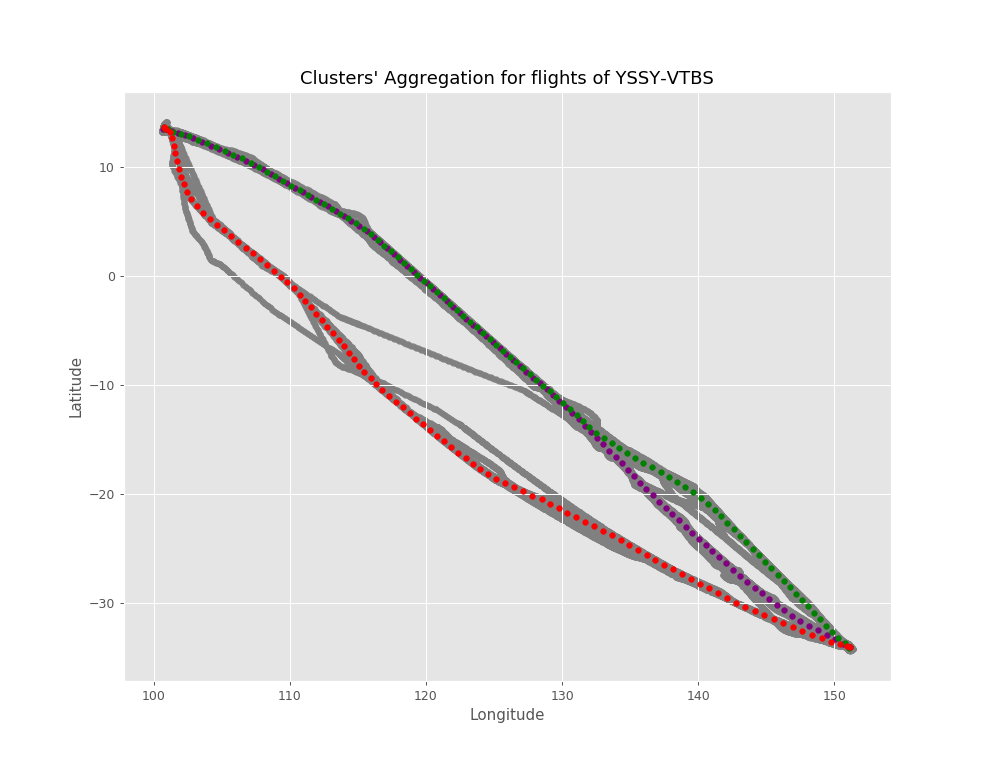} 
  \caption{Aggregated Routes in case of Sydney-Suvarnabhumi airports}
  \label{fig:yssy_vtbs_agg}
  \end{subfigure}\par\medskip
  \caption{Aggregated Routes as a final output of our framework}
\label{fig:agg}
\end{figure}

Last but not least, after clustering the routes into particular clusters, we then employ a simple route matching technique to aggregate all flights' paths in each cluster into unique route that represents for the air route of each pair of airports (see the Figure \ref{fig:agg}). This visualization, once employed in real application, will act as a visual suggestion for Airline Route Planning. In summary, a statistical report of the most important information from our framework are put in the Table \ref{table:our_framework_sum}, taking into account the experiments in three pairs of airports.
\begin{table}
\caption{The Summary of Proposed Framework}
\centering
\label{table:our_framework_sum}
\begin{tabular}{p{2.4cm} c c c}
\toprule
\multicolumn{1}{l}{Pair of Airports} & \multicolumn{1}{l}{NZCH-WSSS} & \multicolumn{1}{l}{YBBN-WSSS} & \multicolumn{1}{l}{YSSY-VTBS}\\
\midrule
No. Flight Path & 43 & 152 & 94\\
\midrule
No. Original Trajectories & 65102 & 169545 & 135586\\
\midrule
No. Interpolated Trajectories & 4300 & 15200 & 9400\\
\midrule
MinPts & 2 & 30 & 3\\
\midrule
Epsilon & 0.523 & 1.583 & 0.603\\
\midrule
Noise Percentage (\%) & 30.00 & 2.98 & 18.87\\
\midrule
No. Clusters via Quantitative Metric & 4 & 2 & 2\\
\midrule
No. Clusters via Visual Inspection & 4 & 2 & 3\\
\bottomrule
\end{tabular}
\end{table}

\section{Conclusions}\label{sec_conclusion}
Based on the potentials and advantages of ADS-B data 
,we propose in this paper a simplified and workable framework for air trajectories clustering which is one of the essential parts in the air management process. To our best knowledge, there is a lack of in-depth analytical study from machine learning based approach, with emphasis on this promising data technology. The experimental results, carried out using the data of three different pairs of airports, can show the effectiveness of our proposed framework, as well as the clustering results based on the combination of two different metrics. Finally, from using the interpolation considered in the framework, it is able to reduce significantly the complexity of data processing, which is very important in real application. The whole framework's source code is made available in github\footnote{\url{https://github.com/quandb/atc}}. In this direction for the future research, we are going to consider the framework's extension to generate predictive capabilities in measuring the operation performance in air traffic management. Furthermore, we also aim to provide guidance and auto-generate inputs for real time decision support system.  


\bibliographystyle{IEEEtran.bst}
\bibliography{IEEEabrv.bib,bibliography.bib}
\end{document}